\documentclass[aip,
rsi,
amsmath,
amssymb,
reprint]{revtex4-1}

\usepackage{graphicx}
\usepackage{dcolumn}
\usepackage{bm}

\usepackage[utf8]{inputenc}
\usepackage[T1]{fontenc}
\usepackage{mathptmx}
\usepackage{etoolbox}

\usepackage{xcolor}

\usepackage{xspace}

\newcommand\NotASection[1]{}

\newcommand\percent[1]{#1\%}

\newcommand\red[1]{#1}

\newcommand\SampleID{\red{VA0605}\xspace}

\newcommand\Temperature{T}

\newcommand\GateDelay{\tau}

\newcommand\SampleTemperature{\red{T_{S}}}
\newcommand\MixingTemperature{\red{T_{S}}}

\newcommand\Window{\red{\tau_{W}}}

\newcommand\meV{~\mathrm{meV}}
\newcommand\eV{~\mathrm{eV}}

\newcommand\FigureReference[1]{\textrm{Figure~\ref{#1}}}

\newcommand\nuRange{0.3\THz\leq \nu\leq 7.0\THz}
\newcommand\BaseTemperature{145\mK} 
\newcommand\PivotTemperature{300\mK}
\newcommand\Bmax{7\Tesla} 
\newcommand\Brange{-\Bmax\leq B\leq +\Bmax} 

\newcommand\MagneticField{B}
\newcommand\MagneticVector{\vec \MagneticField}

\newcommand\refa[1]{\FigureReference{#1}(a)}
\newcommand\refb[1]{\FigureReference{#1}(b)}
\newcommand\refc[1]{\FigureReference{#1}(c)}
\newcommand\refd[1]{\FigureReference{#1}(d)}

\newcommand{\lineitembf}[1]{\item \textbf{#1}:}

\newcommand\Kelvin{~\mathrm{K}}
\newcommand\Tesla{~\mathrm{T}}
\newcommand\ps{~\mathrm{ps}}
\newcommand\second{~\mathrm{ps}}

\newcommand\nm{~\mathrm{nm}}
\newcommand\mm{~\mathrm{mm}}
\newcommand\cm{~\mathrm{cm}}
\newcommand\fs{~\mathrm{fs}}
\newcommand\THz{~\mathrm{THz}}
\newcommand\mK{~\mathrm{mK}}
\newcommand\mJ{~\mathrm{mJ}}

\newcommand\uW{~\mu\mathrm{W}}
\newcommand\kV{~\mathrm{kV}}

\newcommand\citeonline[1]{\onlinecite{#1}}
\newcommand\chem[1]{$\mathrm{#1}$}

\newcommand\ii{\mathfrak{i}}

\newcommand\thatis{i.e.,~}
\newcommand\forexample{e.g.,~}

\makeatletter
\def\@email#1#2{%
 \endgroup
 \patchcmd{\titleblock@produce}
  {\frontmatter@RRAPformat}
  {\frontmatter@RRAPformat{\produce@RRAP{*#1\href{mailto:#2}{#2}}}\frontmatter@RRAPformat}
  {}{}
}%

\makeatother
\begin{document}
\title{Broadband Terahertz Time-domain Spectroscopy of Quantum Materials in a Dilution Refrigerator}

\author{R. J. Vukelich}\thanks{Contributed equally to this manuscript.}
\author{T. Norden}\thanks{Contributed equally to this manuscript.}
\affiliation{Department of Physics, Baylor University, Waco, TX, 76798-7316, USA}
\author{T. G. Hastings}
\affiliation{Department of Physics, University of Alabama at Birmingham, Birmingham, AL, 35294-1170, USA}
\author{M. Giri}
\affiliation{Department of Physics, Baylor University, Waco, TX, 76798-7316, USA}
\author{M. Caldwell}
\affiliation{Department of Physics, Baylor University, Waco, TX, 76798-7316, USA}
\author{S. Forutan}
\affiliation{Department of Physics, Baylor University, Waco, TX, 76798-7316, USA}
\author{J. L. Reno}
\affiliation{Center for Integrated Nanotechnologies, Sandia National Laboratories, Albuquerque, New Mexico 87185, USA}
\author{D. J. Hilton} 
\affiliation{Department of Physics, Baylor University, Waco, TX, 76798-7316, USA}
\email{David\textunderscore Hilton@baylor.edu}

\date{\today}

\begin{abstract} We have constructed a terahertz time domain spectroscopy system using a Blue\-fors dilution refrigerator with a $\Bmax$ split-coil magnet. Using a gallium arsenide single quantum well sample, terahertz wave\-forms were measured at $\BaseTemperature$ in a magnetic field range from 0 to 6 Tesla to measure cyclotron resonance. Effective mass is found to be $0.073 m_{e}$, which is larger than the commonly accepted bulk value of $0.068 m_{e}$.
\end{abstract}
\maketitle

\section{Introduction}
Quantum materials have electronic and optical properties governed by \emph{non\-trivial} quantum mechanical interactions\citep{Cava2021}.   The understanding of the physics of quantum materials is critical for the development of next-generation electronic and optical devices.  In addition to the recent developments in quantum computers\citep{Google:2024gg,Quantum:2025qq}, quantum sensors and other quantum devices also rely on operation at these temperatures and, therefore on the electronic phase of materials near their ground state\citep{Hausmann2014dt,Degen2017,Marciniak:2022mm, Aslam2023, Esat2024}.   

Modern electronics and opto\-electronics are generally dependent on either Group III-V (\forexample \chem{GaAs}\citep{Adachi:1985aa}, \chem{Al_xGa_{1-x}As}\citep{Adachi:1985aa}) or Group IV (\forexample \chem{Si}\citep{Luttinger1955ee}, \chem{SiGe}\citep{Lee2005ll}, \chem{Ge}\citep{Luttinger:1955ll, Luttinger:1956ll}) semiconducting compounds.  These conventional semiconductors and metals still require a quantum mechanical description via band structure theory, but their properties can be largely described within the single-particle approximation\citep{Brandt1970,Adachi:1990aa} or with, at most, weakly interacting systems\citep{Hanke1979} at all but the lowest of temperatures. These are, as a result, generally not classified as quantum materials\citep{Cava2021}.  

The dependence of quantum materials on \emph{non-trivial} interactions and wavefunction interference leads to a wide diversity of emergent behavior.  One of the earliest examples of quantum behavior in materials was the observation of the integer (IQHE)\citep{vonKlitzing1980kw} and later fractional (FQHE)\citep{Tsui1982bv} quantum Hall effect in two-dimensional silicon and gallium arsenide samples.  The IQHE and FQHE result from the formation of edge states in quantized Landau levels due to the interference of electron wavefunctions in the edge states, which results in the formation of off-diagonal conductivity plateaus\citep{Laughlin1981bw}. Other classes of quantum materials include both conventional\citep{BardeenPhysicalReviewLetters1958} and unconventional superconductors\citep{Bednorz1986gk}, Weyl semi\-metals\citep{Yan2016}, and other topological materials\citep{Sushkov2010dd,Jenkins2010kg, Neupane2012ft,Aguilar2012,Luo2019gy}.    An extensive recent review of quantum materials and their applications can be found in ref.~\onlinecite{Zong2023}.

Materials development by serendipity has been a common pathway for many classical and quantum materials\citep{Kalinin2015}.  \emph{Design} of quantum materials is challenging due to their strong dependence on these complex interactions\citep{Basov:2017bb}. A number of empirical and quantitative rules aid in the development of conventional materials\citep{Xie2024}, while the corresponding design rules for quantum materials are less well-developed\citep{Keimer:2017kk, Ahn2021, Giustino2021, Marzari2021}.  The development of novel experimental tools that can elucidate these quantum interactions is critically needed to permit the development of new quantum materials.  

Optics-based experiments provide direct spectroscopic information on the relevant energy scales of these interactions over a wider range of excitation energies than are accessible to transport techniques\citep{Basov:2017bb}.  Ultrafast experimental techniques can complement these measurements by probing the \emph{dynamics} of coupling between different degrees of freedom in quantum materials\citep{Fausti2011dy}.  In this manuscript, we describe the development of a broadband ultrafast terahertz time-domain spectroscopy experiment.  Our experiment has a bandwidth of $\nuRange$ that is generated via frequency mixing of near-infrared femtosecond pulses in a plasma\citep{Kim2008eq, Karpowicz2009jj}.  Terahertz pulses were collected in transmission geometry through our samples and detected using THz-ABCD detection methods\citep{Lu2009ko} at temperatures $\Temperature \geq \BaseTemperature$ and in external magnetic fields ($\Brange$).   To circumvent the limitations of fiber optics at these frequencies, our system employs free-space operation. The terahertz pulses are generated outside the dilution refrigerator, transmitted through the sample chamber via suitable optical windows, and then detected outside the refrigerator. After outlining our design decisions, we demonstrate the functionality of our instrument by studying cyclotron resonance in a Landau-quantized two-dimensional electron gas sample\citep{Wang2007dm, Curtis2016jm, Barman2023}.  These measurements represent the first ultrafast terahertz optical study of materials at sub-$300\mK$\citep{Curtis2016jm}, opening a new experimental regime for the characterization and development of quantum materials. 

\section{Background}
\subsection{Transport-based characterization}
Low-temperature characterization techniques have been extensively employed to study the complex and often competing interactions between electronic, lattice, spin, and orbital degrees of freedom in  materials\citep{vanderPauw1958ui,Meaden1971, Miccoli2015eq, Basov2011ht}. In the case of quantum materials, multiple experimental techniques can be needed to unravel the complex competition between these degrees of freedom\citep{Basov2017}.  The study of materials at temperatures below $\approx 1\Kelvin$ typically requires specialized hardware and sophisticated cooling techniques to provide sufficient cooling power to reach these temperatures\citep{London1962,Hall1966,Wheatley1968}.  Low-temperature measurements of quantum materials have been enabled by the development of the  dilution refrigerator\citep{London1962,Hall1966,Wheatley1968,Hata2013hh,Cao2021}.   Dilution refrigerators exploit the phase separation of mixtures of He-3/He-4 below $T=0.8\Kelvin$ into a surface He-3 rich phase and the remaining He-3 poor phases\citep{Goldstein1954}. He-3 atoms from the rich phase move into the He-3 poor phase and absorb heat (enthalpy) to do so.  This He-3 is extracted from the mixture and recirculated through the system to allow the mixing chamber to reach temperatures below $10\mK$.

Characterization of materials at temperatures below $1\Kelvin$ has traditionally been performed using transport-based techniques.   This allows for the sample chamber to be isolated from ambient conditions without the need for optical windows and reduces the heat load on the sample\citep{stewart1983,Pobell2007}.  There has been significant development of high-sensitivity techniques to measure electrical properties that can maintain this thermal isolation of the sample space that is necessary to reach these low temperatures.  Measurements of specific heat\citep{bachmann1972, stewart1983, wilhelm2004} and resistivity/conductivity are common and the interpretation of these results has enabled the elucidation of the properties of these materials\citep{pellegrino2016}.

\subsection{Optics-based characterization}
Optical spectroscopy of condensed matter systems permits us to measure the absorption, reflection, and transmission of materials across a wide range of photon energies, from the microwave through the x-ray, and over a broad range of time scales, from the femtosecond ($10^{-15}\second$) to time-integrated measurements over an extended time frame.  Each wavelength range is sensitive to different electronic processes in the material that is being studied.  In semiconductors and metals, the free carrier response can be studied using microwave\citep{Brodwin:1965bb} and/or terahertz spectroscopy\citep{Schmuttenmaer:2004jm, LLoyd-Hughes:2012ll}. In superconducting samples, both of these frequency ranges study the Cooper pairs in the superconducting condensate and can be used to determine the pairing energy and quasi\-particle density\citep{Basov:2005bb}. On the other end of the spectrum, the x-ray response results in photoemission of electrons\citep{Nagaoka2025, vanderHeide2011} or by the generation of a diffraction pattern\citep{Nagaoka2025,Waseda2011} that can be used to characterize the crystalline order of materials\citep{FIT2D2016}.  The choice of photon energy range under study determines the materials physics under test.

Optical fibers are one  method of light delivery into the sample chamber\citep{Crooker2002ji,Onyszczak2023, Hamamoto2025RSI} in the absence of windows.    Fiber and other wave\-guide coupling techniques are often used with these low-temperature spectroscopy experiments to enable delivery of visible or near-infrared light into the sample space and to collect the transmitted or reflected light for analysis\citep{Ciesiokiewicz2025}.  At these wavelengths, optical fibers are usually based on fused silica near its dispersion or loss minima\citep{Agrawal2001ts}.     Ultrafast delivery of broadband pulses would be significantly degraded by material dispersion, which will stretch the pulse duration and limit the temporal resolution of the system\citep{Gobel2004dt}.  Nonlinear phenomena such as Brillouin scattering\citep{Kosugi1999, Agrawal2001ts} can complicate pulse propagation, as well, and lead to power transmission limit or unstable frequency spectra due to super\-continuum generation\citep{Gaeta2000iz}.

Our instrument design focuses on generation and detection of terahertz light over the spectrum from $\nuRange$.  Metal waveguides are one method for guided terahertz light.  This couples the terahertz light to the surface plasmon mode of a stainless steel rod for long term transport\citep{Wang2004iu}.  The configuration in ref.~\citeonline{Wang2004iu} has bandwidth that is limited by the source to $0.1\THz$ to $1.0\THz$ and has limited coupling efficiency between its linearly polarized source and the radially polarized ($E \sim \hat{r} E_R$) surface plasmon-polariton mode, which limits its utility for spectroscopic applications.  Other alternatives based on hollow waveguides\citep{Arunkumar2012} and all-polymer waveguides\citep{Wang2011wg} suffer from limited bandwidth and mode dispersion that would make them difficult to use with broadband ultrafast terahertz sources.

Ref.~\onlinecite{Onyszczak2023} recently demonstrated Fourier Transform Infrared (FTIR) spectroscopy measurements in a dilution refrigerator at temperatures as low as $43\mK$.   FTIR is a highly flexible platform that is capable of operating over a wide range of the electromagnetic spectrum if a set of suitable light sources and beam splitters are available. This method is able to provide important information about a wide range of optical conductivities for materials in equilibrium.  Their system employed either a globar or a mercury arc lamp as the incoherent light source from $0.25\THz$ ($h\nu = 1\meV$) to $250\THz$ ($h\nu = 1\eV$).  This light is then coupled via free-space optical windows into the dilution refrigerator.  While FTIR has been previously demonstrated to have broad utility\citep{Basov2017} to characterize quantum materials, approaches based on \emph{coherent} sources in the far infrared and terahertz have been shown to have advantages over incoherent sources for high mobility materials\citep{Chou1988fd}.

THz spectroscopy using a hollow metal waveguide in a dilution refrigerator has been previously reported\citep{Vaughan2022}.  Here, the THz was generated using a quantum cascade laser (QCL)  inside of a dilution refrigerator with a central frequency of $2.6\THz$. They demonstrated an operating temperature of $160\mK$ at the mixing chamber and an electron temperature at the sample of $\sim 430\mK$ while measuring cyclotron resonance.  This technique has recently been modified to be done in free space within the dilution refrigerator and has seen a significant reduction in thermal loading, with an operating temperature of $47\mK$ \citep{Vaughan2025}. Potential limitations with this method include the fixed narrow band output commonly seen in QCL's, which does not easily allow for the broadband experiments that we will demonstrate in this manuscript.

\subsection{Ultrafast optics-based characterization}
\emph{Ultrafast} optical spectroscopy techniques permit us to study these materials far from equilibrium.  We can use these to determine how the system returns to equilibrium and the time-scales on which it does so.  Ultrafast sources typically use broadband, sub\-picosecond pulsed lasers in a pump-probe \citep{Prasankumar:2011pp}, transient four-wave mixing\citep{Bristow:2009bb}, time-resolved photo\-luminescence\citep{Achermann:2016aa} or other experimental configuration to study the nonlinear susceptibility of materials\citep{Boyd:1984bb}.  These experiments use broadband sources and high peak amplitudes that often exclude the potential of using dispersive wave\-guides that will necessarily compromise the time-resolution of the experiment\citep{Ouzounov:2003oo,Chong:2007cc, Phoenix:2020pp}.  

\begin{figure}
    \centering
    \includegraphics[width=1.0\linewidth]{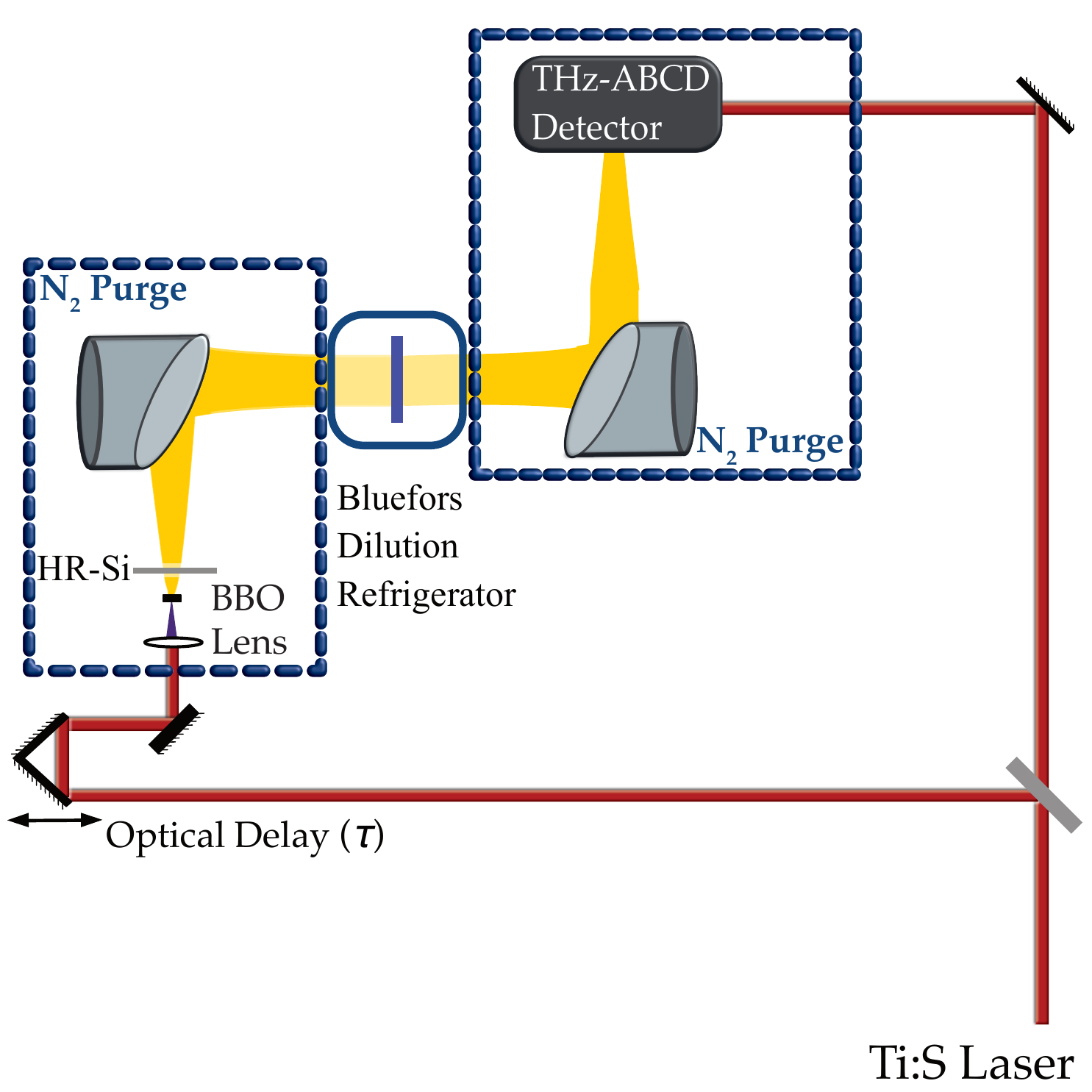}
    \caption{We generate terahertz pulses by mixing the fundamental [$\lambda_1 = 800\nm$] and second harmonic [$\lambda_2 = 400\nm$, generated using \chem{\beta-BaB_2 O_4} (BBO)] of the titanium:sapphire output in a plasma filament in air\cite{Kim2008eq}.  Terahertz emission is separated from the residual visible light using a high-resistivity silicon filter (HR-Si).   This is then collected by an off axis parabolic mirror into the Bluefors dilution refrigerator.  Optical windows in and out of the sample chamber are TPX (polymethylpentene), which has adequate transmission in both the visible and over our terahertz bandwidth\citep{Bichon2022}.  After the dilution refrigerator windows, the transmitted light is incident on a custom-constructed THz-ABCD detector\citep{Ho2010jv}, which we use to recover the electric field, $\vec E\bigl(\GateDelay\bigr)$, of the transmitted THz pulse. }
    \label{fig:TTDS}
\end{figure}

\section{Instrument Design}
A diagram of our experimental setup is shown in \FigureReference{fig:TTDS}.  We use an ultrafast amplified titanium:sapphire laser system to generate linearly-polarized broadband ($\nuRange$) terahertz pulses by mixing the fundamental ($\lambda_1 = 800 \nm$) and second harmonic ($\lambda_2 = 400\nm$) in a plasma\citep{Kim2008eq, Karpowicz2009jj}.  We generate the plasma at the focus on the titanium:sapphire laser beam in nitrogen and generate the second harmonic of the beam using a \chem{\beta-BaB_2O_4} single crystal.  To block the residual $800\nm$ and $400\nm$ light, we use a $1\cm$ thick high resistivity silicon window as a visible light filter to block all but the generated terahertz light from transmission into our dilution refrigerator.    We use a pair of off-axis parabolic mirrors to guide the terahertz beam from the plasma filament into the sample chamber and to collect the transmitted beam and bring it to our THz-ABCD detector\citep{Ho2010jv}.  

\begin{figure}
    \centering
    \includegraphics[width=0.85\linewidth]{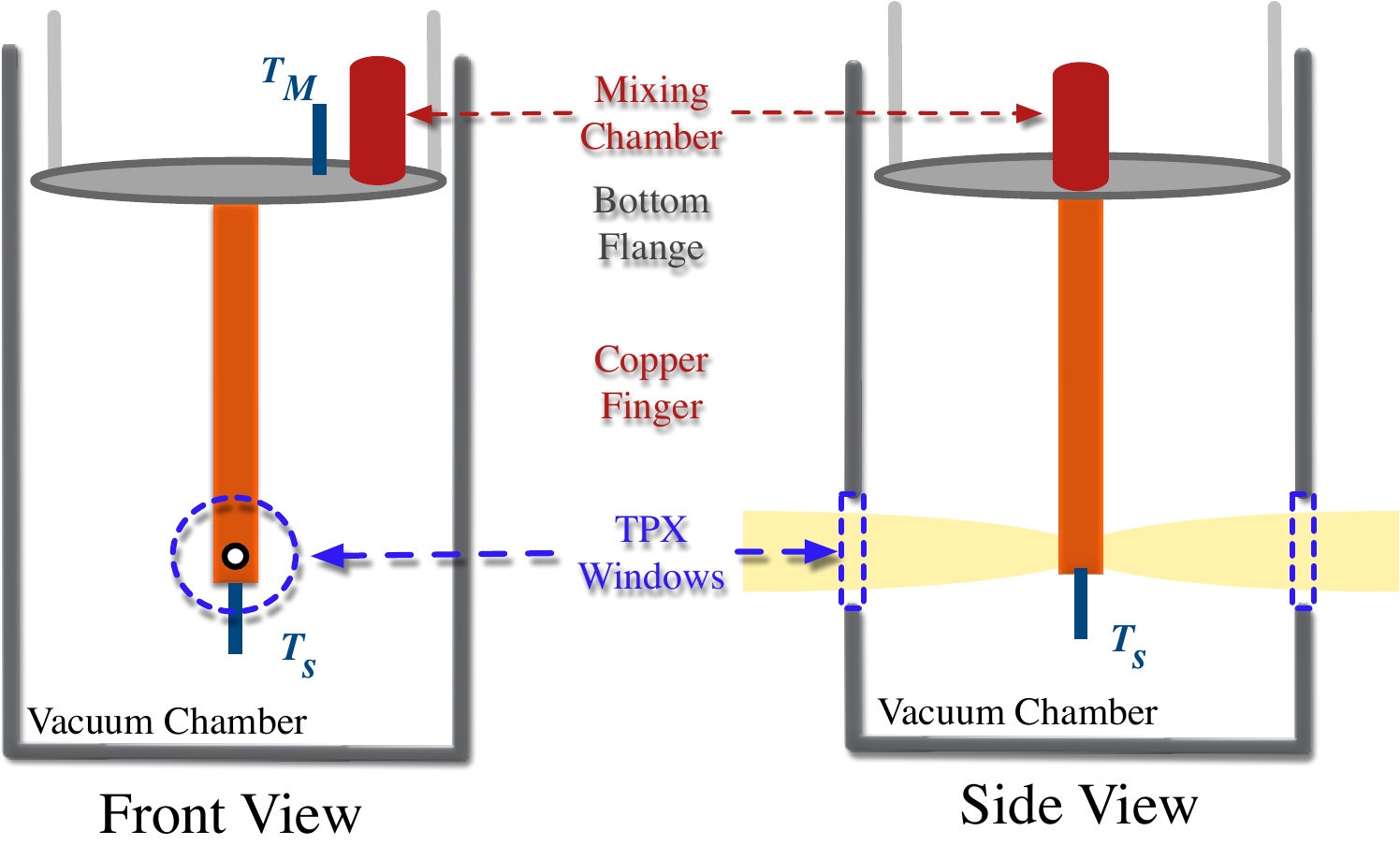}
    \caption{Inside the dilution refrigerator, the lowest flange has the He-3/He-4 mixing chamber along with a co-located temperature sensor ($\MixingTemperature$).  We have custom constructed a copper cold finger that is in thermal contact via an indium foil seal with the lowest flange.  The cold finger has a $1\cm$ transmission aperture that is aligned with the TPX optical windows.  A second temperature sensor (RuOx Sensor, Model number: A01892, $\SampleTemperature$) is attached to the copper finger at the sample position and will be the reported sample temperature in this manuscript.  Samples are mounted on the transmission aperture using GE 7031 varnish, which ensures good thermal contact between the cold finger and sample \citep{Stephens1971}. (Not shown) There are three internal radiation shields with a pair of windows each.  This results in eight TPX windows on the full beam path.}  
    \label{fig:chamber}
\end{figure}

A simplified diagram of the dilution refrigerator chamber is included as \FigureReference{fig:chamber}.  In the Bluefors system, the bottom flange has the He-3/He-4 mixing chamber and a temperature sensor ($\MixingTemperature$).  This is typically $\MixingTemperature \approx 10\mK$ at the base temperature of our experiment.  We mount our sample on a custom-constructed copper finger that is in good thermal contact via an indium foil seal with the bottom flange.  This custom copper finger has a $d=1\cm$ hole on which we mount our samples to measure the transmitted terahertz light.  This design positions the sample in the terahertz beam path through the TPX windows while maintaining thermal contact with the mixing chamber.  A second temperature sensor ($\SampleTemperature$) is located on the bottom of our copper finger close to the sample; this is the sensor recording used for all temperature readings in the experiments described in this manuscript.

\subsection{Design Criteria}
Specific design criteria for our apparatus include:
\renewcommand{\theenumi}{\Roman{enumi}.}
\begin{enumerate}
\lineitembf{Broadband THz generation and detection} We use a broadband ultrafast terahertz spectrometer similar to the one described in ref.~\onlinecite{Curtis2018}.  This generates terahertz by frequency mixing of the fundamental and second harmonic of a titanium:sapphire laser in air\citep{Karpowicz:2009kk}.  We detect these using the THz-ABCD detector geometry described in ref.~\onlinecite{Lu:2009ll}. This system will have a broad bandwidth to achieve sub\-picosecond time-resolution or, equivalently, the bandwidth will span from 0.3 THz ($h\nu = 1.2\meV$) to the edge of the terahertz transmission window in our TPX windows, which is beyond $7\THz$ ($h\nu = 29.0\meV$)\citep{Podzorov:2008pp}.
\lineitembf{Low-temperature operation in the quantum limit} This experiment will allow for the study of materials using ultrafast terahertz sources at temperatures below $\SampleTemperature\leq 200~\mathrm{mK}$ ($k_B \SampleTemperature \leq 17.2~\mu\mathrm{eV}$).  Normalized energy scales, $x = \frac{h\nu}{k_B \SampleTemperature}$ in this geometry can be as large as $x \approx 1,700$, demonstrating the ability of our setup to probe samples in the quantum limit where thermal fluctuations are minimized.  To do so, we use external generation and detection methods to minimize internal heat load in the sample chamber\citep{Keimer:2017kk}.  We also restrict the solid angle into the sample chamber to minimize the effects of stray light into the sample position.  When we replace the optical windows with metallic plugs, the sample temperature is below $\MixingTemperature\leq 6\mK$.  Thus, the base temperature in our experiment is limited by the transmission of the optical windows and our success in isolating the sample chamber from the ambient.  Proper thermal management of ambient radiation into the sample chamber is an area of future system optimization to reduce our base sample temperature below the current $\SampleTemperature= \BaseTemperature$ to extend this deeper into the quantum limit (\thatis larger $x$).
\lineitembf{External magnetic field} Our system includes an integrated $\Bmax$ split-coil magnet to enable us to study materials at low temperature and external magnetic field in both the Faraday ($\MagneticVector \parallel \vec k$) and Voigt geometries ($\MagneticVector \perp \vec k$), where $\vec k$ is the propagation vector of the terahertz field.  This is below the critical magnetic fields for several superconducting systems of interest (\forexample $B_{c2} \approx 15\mbox{~to~}20\Tesla$ in many cuprates\citep{Basov:2005bb} and even higher in iron-based superconductors\citep{Paglione:2010pp, Keimer:2017kk, Fernandes:2021ff}), which will require specialized facilities to study\citep{Toth2012fha, Curtis:2014di}. 
\lineitembf{Coherent control} The use of ultrafast optical sources in our design instead of black\-body-type sources of terahertz radiation is, in part, motivated by their ability to control the electronic wave function\cite{Arikawa:2011aa} and potentially access hidden order in materials\citep{Wei:2024ww}.  In this context, hidden order emerges when the external stimulus significantly contributes to the equation of state of the system and results in the emergence of a novel electronic, structural, or magnetic phase not otherwise present in the absence of this stimulus.  Recently developed \emph{table\-top} ultrashort terahertz emission sources\citep{Hebling:1996hh,Hebling:2010cf, Kim:2008kk} are capable of non\-perturbatively exciting the sample with picosecond pulses with electric fields on the same order of magnitude or larger than the internal mean fields\citep{Aoki:2014aa,Yang:2023yy}. While this  manuscript will focus on single pulse spectroscopy, straightforward extensions to our configuration will permit coherent control experiments in the future\citep{Arikawa2011cl, ElsaesserNewJournalofPhysics2013}.
 \end{enumerate}

\begin{figure*}[!t]
\includegraphics[scale = 1.0]{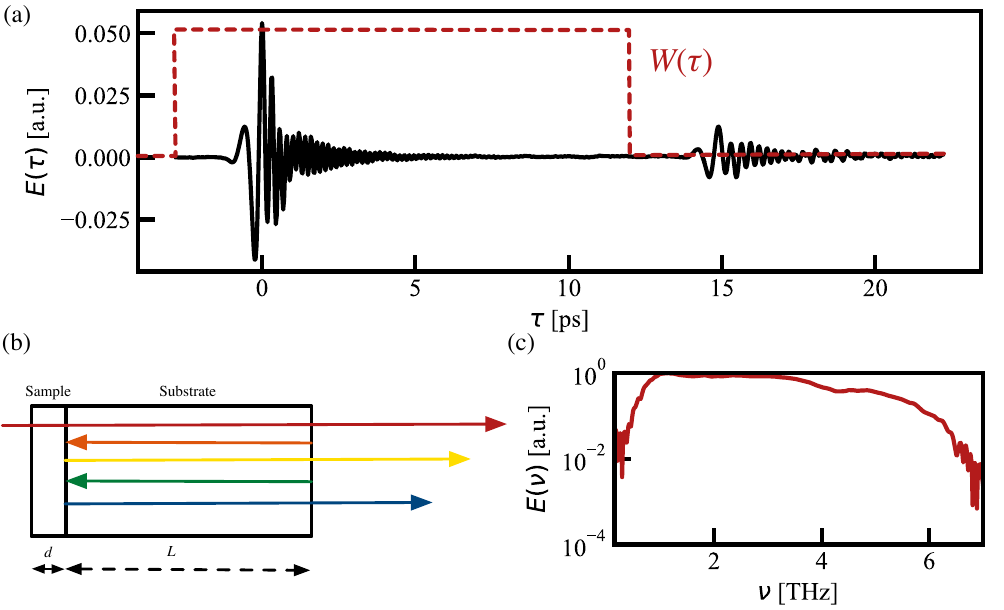}
\caption{\label{fig:Figure_1}(a) A representative THz Waveform, $E\bigl(\GateDelay\bigr)$, and the numerical window, $W(\GateDelay)$ with a width of $\Window = 15\ps$ applied to our data to remove secondary pulse reflections.  (b) This is a simplified diagram of our sample (Sample ID: \SampleID), consisting of the two dimensional layer ($d$) and a thick GaAs substrate ($L$).  The windowed $E_W\bigl(\GateDelay\bigr)$ is the single pass waveform through the structure and removes the multiple reflections/etalon from the calculated spectrum.   (b) A schematic diagram of the sample, showing the origin of the multiple reflections in the time-domain or, equivalently, of the Fabry-Perot interference fringes in the frequency domain.  (c) The FFT of numerically windowed spectrum from the first pass through the sample. }
\end{figure*}

\subsection{System Components}
Our system consists of four main components: a laser source, a terahertz generator, a terahertz detector, and a free-space optics-compatible dilution refrigerator.  The use of free-space optical techniques simplifies the experiment design as it allows for both the generation and detection \emph{outside} of the cryostat.   

 \renewcommand{\theenumi}{\Alph{enumi}.}
\begin{enumerate}
    \lineitembf{\label{sec:TTDS}Titanium:sapphire laser amplifier} The system uses a femtosecond laser regenerative amplifier [Spectra Physics Solstice Ace, $\sigma = 35\fs$, $\mathcal{U} = 7\mJ$ per pulse, $\lambda_0 = 800\nm$ ($h\nu_0 = 1.55 \eV$)] to generate terahertz pulses.  A \percent{90} reflective beam\-splitter is used for the terahertz generation path and gate path, respectively. To generate terahertz pulses for both Faraday and Voigt geometries, this generation pulse is split by a 50:50  beam\-splitter, resulting in $3.15\mJ$ pulses that are used for both geometries. In this paper, we focus on the Faraday geometry generation path used to observe cyclotron resonances in a gallium arsenide quantum well sample.
    \lineitembf{Terahertz time domain spectroscopy via plasma generation\label{sec:plasma}} Terahertz pulses are generated using the method of gas-ionization. The fundamental pulse, $\nu_0$, of the Ti:S is focused through a \chem{\beta-Ba_2BO_4} ($\beta$-barium bor\-ate) Type-I cut crystal in a nitrogen purged box, resulting in a collinear second harmonic pulse at $2\nu_0$. A photo\-current is induced in the ionized nitrogen as the electrons move away from the ions when there is a phase difference between the $\omega$ and $2\omega$ photons at the focus \cite{Kim2007oe, Kim:2008kk}. The result of this induced photo\-current is the generation of ultra\-fast terahertz radiation with $\sim 1\uW$ of average power, along with a white light continuum, that is reflected off an off-axis parabolic. A high resistivity silicon filter is used to separate the broadband terahertz frequencies from the high frequency components of the white light generation and the residual titanium:sapphire beam.
    \lineitembf{\label{sec:ABCD}Air-based coherent detection} An air-based coherent detector is used to detect the terahertz pulses \citep{Lu2014ll}. After transmission through the dilution refrigerator, the terahertz pulse is focused by an off-axis parabolic between two copper electrodes. These copper electrodes are spaced apart $1.5\mm$ and connected to an Advanced Energy Ultra\-volt $2\kV$ bipolar high voltage amplifier, resulting in a modulated bias field within the capacitive gap. The biased field is modulated using a custom circuit creating a bipolar signal modulated at half the frequency of the titanium:sapphire amplifier’s repetition rate. This allows the bipolar biased field to simulate a quasi-second-order nonlinear process, allowing for the coherent detection of the terahertz pulse by using a Newport Femto\-watt Detector to detect the filtered 400 nano\-meter generation created by the terahertz pulse, gate pulse, and biased field mixing. This signal is sent to a lock-in which is locked to half the titanium:sapphire amplifier’s repetition rate.
    \lineitembf{\label{sec:DR}Dilution refrigerator} A Bluefors LD-400 dilution refrigerator is used to cool samples down to the mK range. The dilution refrigerator is a cryogen-free system that has two perpendicular optical axes through an external shield (room temperature), radiation shield ($50\Kelvin$), $\Bmax$ cryogen-free horizontal field split-coil magnet (American Magnetics, Inc.) that can be replaced by a second radiation shield ($4\Kelvin$), and still tail ($800\mK$). A He-4 compressor in a closed system is used to initially cool down to liquid helium temperatures, where a He-3/He-4 mixture can be introduced to be pre\-cooled. After the mixture is pre\-cooled, it can be condensed to a state where He-4 is pulled to the bottom of the dilution chamber by gravity and the He-3 is forced through the condensed He-4. The resulting osmotic pressure cools the dilution chamber down to mK temperature, which is connected to the mixing chamber flange.  Available optical axes allow us to conduct measurements in either Faraday or Voigt geometries when a sample is in the dilution refrigerator. Each optical axis consists of four optical windows before and after the sample, for a total of eight optical apertures (not shown in \FigureReference{fig:chamber}). For this experiment, TPX poly\-methyl\-pentene windows were used in the Faraday axis optical apertures. 
    \lineitembf{\label{sec:mount}Sample mount} A custom copper cold finger with optical aperture has been connected to the mixing chamber flange and extends down so the sample mount’s aperture is aligned with the optical axis of the external shield windows. GE Varnish is used to mount samples to the cold finger and copper tape is used to ensure there are no gaps between the sample and cold finger aperture through which terahertz radiation is transmitted. To provide an accurate temperature measure at the sample position, a second temperature sensor is connected to the bottom of the cold finger in proximity to the sample. The cold finger can be rotated 90 degrees to allow for measurements in Faraday ($\vec k \perp \vec B$) or Voigt ($\vec k \parallel \vec B$) geometries.
    \lineitembf{Cryostat windows\label{sec:window}} Our broadband generation is limited by the transmission of the windows used in this cryostat.   A wide range of different materials can be used for this frequency range, including diamond, sapphire, silicon, and quartz as well as multiple plastics including polyethylene, polymethylpentene (TPX), and Polytetrafluoroethylene (Teflon)\citep{Schmuttenmaer2004jm, JLY2012}.  We have chosen $1\mm$ thick TPX windows for our system, which is motivated by its high transmission in both the terahertz as well as in the visible/near infrared\citep{Curtis2014di}.  
\end{enumerate}

\section{Results and Discussion}
We use our apparatus to study cyclotron resonance in a high mobility two-dimensional electron gas (2DEG).   Recent work has emphasized new 2D materials based on monolayer semiconductors\citep{Bolotin2009ko, Vogt2012cf, Tao2015hq, Davila2014bs, Zhu2015hc} as well as monolayer transition metal dichalcogenides.  This has also extended into multilayer systems with controlled twist angles to directly engineer the electronic and optical properties of these materials\citep{Carr2017}.  While these provide significant methods for designing and controlling the electronic properties of the 2DEG, growth of high mobility samples that are predicted to have the longest coherence lifetimes are enabled by high-purity III-V molecular beam epitaxy\citep{Manfra2014dj}.  High mobility two-dimensional systems are known to have narrow band and magnetically tunable resonances that will permit us to examine the resolution limits of our system and plan for future upgrades and motivate the development of new analysis methods for future use.

Our sample is a gallium arsenide quantum well (Sample ID: \SampleID) doped to an electron concentration of $n_s \approx 4\times 10^{11}~\mathrm{cm}^{-2}$ that is grown via molecular beam epitaxy\citep{Stormer1979ge}.  In the presence of an external magnetic field, the states split into a discrete spectrum of equally spaced Landau levels, $E_n = n \hbar \omega_{CR}$, separated by the cyclotron energy.  These Landau levels are broadened by sample-dependent disorder that results from lattice impurities and defects in the material\citep{Barman2023}.   These levels are also broadened by the collective emission of this ensemble of dipoles as super\-radiant emission\citep{Jho2006cp}, which makes the decay time result from both phenomena, depending on the temperature range\citep{Barman2023}. These Landau levels result in a strong circular dichroism near the cyclotron energy that results from a transition between the highest filled and lowest unfilled Landau levels with dephasing time determined by the sample disorder\citep{Barman2023}.

\begin{figure*}
\includegraphics[scale=1.0]{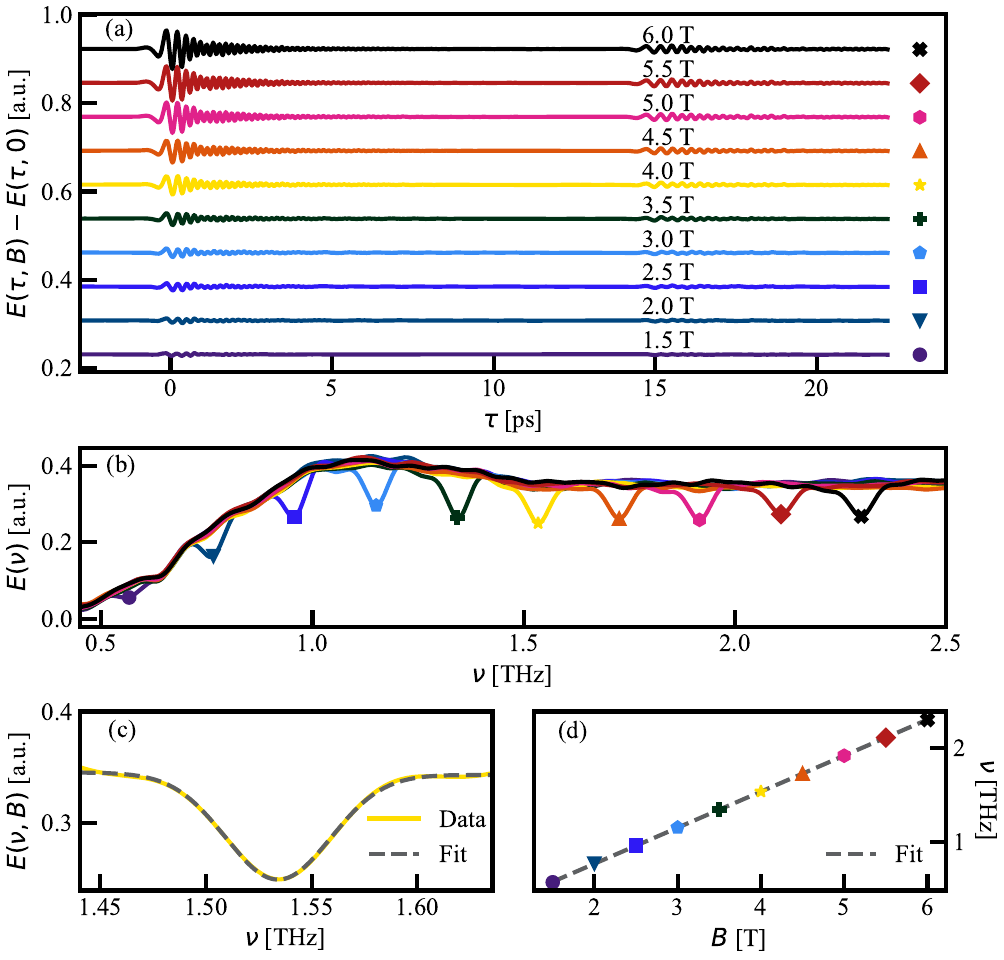}
\caption{\label{fig:Figure_2} (a) The change to the transmitted terahertz waveforms at $B$ when compared to the zero field transmitted THz waveform at $\BaseTemperature$ for $1.5\Tesla$ to $6.0\Tesla$.  For fields below $1.5\Tesla$ and above $6.0\Tesla$, the cyclotron resonance in GaAs is at a frequency outside of our experiments spectral bandwidth ($\nuRange$).  (b) Frequency spectra of the measured wave\-forms calculated by taking the fast Fourier transform of the waveform data in (a).   (c) A representative fitting of the spectrum line shape to a homogeneously broadened Lorentzian line shape\citep{Hilton2012}.  (d) The central frequency, $\nu_c\bigl(B\bigr)$ is shown on as a function of $B$.  We determine the effective mass from the slope of this fitting, as discussed within the text and in Ref~\citeonline{Wang2007dm}.  The line width, $\Delta \nu$ is instrument-response limited in our apparatus by the TPX window thickness and will be the focus of future development efforts.}
\end{figure*}

The sample temperature was measured to be $\SampleTemperature= 141\mK$ without the incident THz beam incident on the sample.  With the THz incident, the base temperature at the sample location rose to $\SampleTemperature=\BaseTemperature$, indicating there is minimal heating from the laser source. Terahertz time-domain spectroscopy wave\-forms were measured at $\SampleTemperature=\BaseTemperature$ in a gallium ar\-se\-nide 18 nano\-meter single quantum well (\SampleID) for magnetic fields from 0 to $6\Tesla$ in $0.5\Tesla$ steps. A waveform measured with no magnetic field is shown in \refa{fig:Figure_1}. The waveform shows the main terahertz pulse and the first satellite pulse, due to internal reflections within the gallium ar\-se\-nide substrate. An example of the internal reflections and the satellite pulses produced is shown in \refb{fig:Figure_1}. The separation between the main pulse and satellite pulse is similar to that found by Ref.~\onlinecite{Curtis2018} of approximately $\Window = 15\ps$ and is determined by the substrate thickness on which this sample is grown.

We numerically window the acquired THz pulse $E_W\bigl(\GateDelay) = W(\GateDelay)\times E\bigl(\GateDelay\bigr)$ with a window length of $\Window=15\ps$ to remove the etalon from our TPX windows.   We calculate the complex spectrum, $\tilde E\bigl(\nu\bigr) = E\bigl(\nu\bigr) \exp\bigl(\ii \varphi\bigr)$ of our terahertz pulses using the fast-Fourier transform of the windowed terahertz wave\-form measured without a magnetic field.  The magnitude of the spectrum, $E\bigl(\nu\bigr)$, is shown in \refc{fig:Figure_1}; the phase, $\varphi$, is not shown. The broadband spectrum of the wave\-form is cut\-off at $7\THz$ due to the eight TPX windows plus sample that the wave\-form needs to transmit through the dilution refrigerator.  The low frequency cutoff is limited by diffraction\citep{You1997}.

To demonstrate the system with the $\Bmax$ magnet, cyclotron resonance in our sample is measured from $0.5\Tesla$ to $6\Tesla$; we do not reach the full $\Bmax$ in this experiment as the cyclotron resonance frequency in gallium arsenide is above the usable bandwidth of our pulse at those fields. The resulting cyclotron resonance wave\-forms are shown in \refa{fig:Figure_2}. The isolated cyclotron resonance is found by subtracting the zero magnetic field wave\-form, $E(\GateDelay, 0)$, from the wave\-forms at the $B$-field strengths labeled on the right side in \refa{fig:Figure_2}, $E(\GateDelay, B)$, to generate $\Delta E\bigl(\GateDelay, B) = E\bigl(\GateDelay, B\bigr) - E\bigl(\GateDelay, 0)$.  To obtain the values of the cyclotron resonances, we compute the fast-Fourier transforms of  $\Delta E\bigl(\GateDelay, B)$ measured from 1.5 to $6\Tesla$.  These are shown in \refb{fig:Figure_2} as a function of the frequency for each value of $B$. 

To determine the effective mass of the gallium ar\-se\-nide single quantum well sample, the cyclotron frequency, $\nu_{c}$, of each curve in \refb{fig:Figure_2} is fitted using a Gaussian curve plus a linear component, where the center of the Gaussian fit is taken to be the cyclotron frequency. An example fitting is shown in \refc{fig:Figure_2}, where the value of the cyclotron frequency is $\nu_{c} = 1.53\THz$ from these fittings at $B=4.0\Tesla$. The cyclotron frequency all values of the magnetic field are shown in \refd{fig:Figure_2}. Using these cyclotron frequencies, a linear fit was used to determine the effective mass of the sample which was found to be 0.073 $m_{e}$, where $m_{e}$ is the electron mass. This effective mass is larger than was reported in a similar experiment in  Ref.~\onlinecite{Barman2023}. This may be related to the stronger confinement effects in this sample and will be the subject of a future investigation by our group\citep{Sigg1985bt}.

\section{Conclusions and future directions}
Broadband terahertz time domain spectroscopy was performed using a dilution refrigerator with a $\Bmax$ magnet to measure the cyclotron resonance of gallium arsenide single quantum wells below $\SampleTemperature=\PivotTemperature$. The zero magnetic field waveform and its fast-Fourier transform verifies our capability to transmit a broadband terahertz waveform through the dilution refrigerator and sample, and perform a measurement on the opposite side of the dilution refrigeration than the generation. Cyclotron resonance measurements on a gallium arsenide single quantum well were performed to verify the use of the $\Bmax$ magnet with the broadband terahertz at $\SampleTemperature=\BaseTemperature$. 

Our ability to perform broadband terahertz spectroscopy in a dilution refrigerator, with or without the $\Bmax$ magnet, will allow us to probe non\-equilibrium dynamics of quantum materials in a temperature regime previously inaccessible to broadband terahertz systems. By generating the terahertz external to the dilution refrigerator, the portion of the titanium:sapphire laser not used for terahertz generation or detection could be utilized in other ways, such as a second terahertz generation path to perform two dimensional terahertz spectroscopy or as a pump beam using either the fundamental or a harmonic. These additional uses would strengthen the ability of this system to understand the exotic quantum ground states seen at mK temperatures.

\section*{Conflict of Interests}
The authors have no conflicts to disclose.

\section*{Author Contributions}
\noindent \textbf{R. J. Vukelich}: Investigation (equal), Formal Analysis (co-lead), Visualization (lead), Software (supporting), Writing/Original Draft Preparation (co-lead); \textbf{T. Norden}: Investigation (equal), Formal Analysis (co-lead), Software (lead); \textbf{T. G. Hastings}: Investigation (supporting), Writing/Review \& Editing (supporting); \textbf{M. Giri}: Investigation (supporting), Writing/Review \& Editing (supporting); \textbf{M. Caldwell}: Investigation (supporting); \textbf{S. Fourtan}: Writing/Review \& Editing (supporting); \textbf{D. J. Hilton}: Conceptualization (lead), Funding Acquisition, Project Administration (lead), Supervision (lead), Visualization (supporting), Formal Analysis (supporting), Writing/Original Draft Preparation (co-lead),  Writing/Review \& Editing (co-lead)

\begin{acknowledgments}
THz Detector Development was supported by the National Science Foundation under grant DMR-1919944.  This work was performed, in part, at the Center for Integrated Nanotechnologies, an Office of Science User Facility operated for the U.S. Department of Energy (DOE) Office of Science by Los Alamos National Laboratory (Contract 89233218CNA000001) and Sandia National Laboratories (Contract DE-NA-0003525).
\end{acknowledgments}

\section*{Data Availability Statement}
The data that support the findings of this study are openly available in the \emph{figshare} repository at http://doi.org/10.6084/m9.figshare.31075618, reference number [10.6084/m9.figshare.31075618]. 

\end{document}